\UseRawInputEncoding

\documentclass[10pt,conference]{IEEEtran}
\IEEEoverridecommandlockouts
\usepackage{amsmath,amssymb,amsfonts}
\usepackage{algorithmic}
\usepackage[linesnumbered, ruled]{algorithm2e}
\usepackage{graphicx}
\usepackage{textcomp}
\usepackage{multirow}
\usepackage{diagbox}
\usepackage[table,xcdraw]{xcolor}
\usepackage{array}
\usepackage{hhline}
\usepackage{svg}
\usepackage{subcaption}
\usepackage{caption}
\usepackage{hyperref}
\hypersetup{nolinks=true}
\newcommand{\noteMK}[1]{\textcolor{violet}{[{\bf #1}]}}

\newcolumntype{L}[1]{>{\raggedright\arraybackslash}p{#1}}

\def\BibTeX{{\rm B\kern-.05em{\sc i\kern-.025em b}\kern-.08em
    T\kern-.1667em\lower.7ex\hbox{E}\kern-.125emX}}
\begin{document}

\title{
Deep Dict: Deep Learning-based Lossy Time Series Compressor for IoT Data
}

\author{
\IEEEauthorblockN{Jinxin Liu$^1$, Petar Djukic$^2$, Michel Kulhandjian$^1$, Burak Kantarci$^1$}
\IEEEauthorblockA{\textit{$^1$School of Electrical Engineering and Computer Science} \\
\textit{University of Ottawa, Ottawa, ON, Canada}\\
\textit{$^2$ Nokia Bell Labs}\\
$^1$\{jliu367, mkulhand, burak.kantarci\}@uottawa.ca,~$^2$petar.djukic@nokia-bell-labs.com}
}

\maketitle

\begin{abstract}
We propose Deep Dict, a deep learning-based lossy time series compressor designed to achieve a high compression ratio while maintaining decompression error within a predefined range. Deep Dict incorporates two essential components: the Bernoulli transformer autoencoder (BTAE) and a distortion constraint. BTAE extracts Bernoulli representations from time series data, reducing the size of the representations compared to conventional autoencoders. The distortion constraint limits the prediction error of BTAE to the desired range. Moreover, in order to address the limitations of common regression losses such as L1/L2, we introduce a novel loss function called quantized entropy loss (QEL). QEL takes into account the specific characteristics of the problem, enhancing robustness to outliers and alleviating optimization challenges. Our evaluation of Deep Dict across ten diverse time series datasets from various domains reveals that Deep Dict outperforms state-of-the-art lossy compressors in terms of compression ratio by a significant margin by up to $53.66\%$.

\end{abstract}


\section{Introduction}
Internet of Things (IoT) is a paradigm that connects real-world objects and collects real-time information/data using various sensors, such as accelerometers, gyroscopes, and temperature sensors \cite{stoyanova_survey_2020}. In recent decades, IoT has been widely used in a variety of applications, including smart healthcare, smart homes, connected vehicles, and wearable devices \cite{nauman_multimedia_2020}. In these applications, massive amounts of time series data are created, stored, and communicated as a result of the widespread use of smart devices, industrial processes, IoT networks, and scientific research \cite{wong_recurrent_2018}. However, transmitting such a large amount of time series can be costly in terms of network bandwidth and storage space \cite{Buddhika.2021}; consequently, many studies focus on compressing time series data with a high compression ratio \cite{jensen_time_2017}. Data compression can be roughly classified into two categories: lossless and lossy compression \cite{chiarot_time_2021}. 
As lossy compression introduces errors to decompressed time series, it is important for lossy compressors to contain error-bound or distortion-constraint mechanisms to make a compromise between compression ratio and decompression errors \cite{jin_deepsz_2019}.

Autoencoder (AE) \cite{bank_autoencoders_2021} is one of the important lossy time series compression techniques. An AE encodes time series data as latent states with real values and decodes them as prediction. Compressed latent states are one of the major overheads of compressed data. This work addresses the potential of encoding time series into Bernoulli distributed latent states rather than real-value latent states in order to drastically reduce the size of latent states and improve compression rate. In addition to AE, prediction-based compressors typically employ regression losses such as L1 and L2 \cite{chandak_lfzip_2020}. Given that outliers and noise can significantly impact traditional regression loss functions like L1 and L2, which may not always align effectively with the underlying objective, this research introduces a novel loss function inspired by the principles of entropy coding. This approach aims to provide a more accurate and precise definition of the problem at hand.


This work proposes a new lossy time series data compression technique, namely Deep Dict in order to improve the compression ratio. The contribution of this paper is threefold: 
\begin{itemize}
    \item A new compression framework, called Deep Dict, is proposed for lossy compression of time series data generated from IoT devices such as gyroscopes, and the results demonstrate that Deep Dict achieves a higher compression ratio than state-of-the-art compressors.
    \item This work proposes a novel Bernoulli transformer-based autoencoder (BTAE) that can effectively reduce the size of latent states and reconstruct IoT time series from the Bernoulli latent states.
    \item We introduce a novel loss function called quantized entropy loss (QEL), tailored to the specific characteristics of the problem. QEL surpasses conventional regression loss functions like L1 and L2 in terms of compression ratio performance. This loss function is adaptable for use with any prediction-based compression method that employs uniform quantization and an entropy coder.
\end{itemize}

The rest of the paper is organized as follows. Section \ref{sec: related work} introduces the state-of-the-art lossy data compressors. Section \ref{sec: method} presents the problem statement, the proposed Deep Dict framework, and the details of network architecture. The datasets and comprehensive experiments are described in Section \ref{sec: exp}. Finally, Section \ref{sec: conclusion} presents conclusions along with future directions.

\section{Related Work}
\label{sec: related work}
In the majority of IoT applications, due to resource constraints on the devices, substantial amounts of data are typically offloaded either to edge nodes or to the cloud for the purposes of analytics or decision support \cite{Kumar.2022}. 

Liang \textit{et al.} \cite{liang_error-controlled_2018} propose SZ2 a framework for adaptive prediction-based compression with Lorenzo or linear regression as the predictor. In addition to compressing data with a specified error constraint, SZ2 maximizes peak single-to-noise ratio (PSNR) to ensure the integrity of recovered data with a high compression ratio. 

LFZip is a lossy compressor, proposed by  Chandak \textit{et al.} \cite{chandak_lfzip_2020}, that utilizes machine learning for prediction, quantization, and entropy coding. The auto-regressive predictor of LFZip has two types: normalized least mean square predictor (NLMS) and bidirectional GRU for learning nonlinear patterns in time series.
In LFZip, L2 serves as the loss function, and the maximum absolute error (MAE) is used for distortion measurement. 

Based on SZ2, Zhao \textit{et al.} \cite{zhao_optimizing_2021} offer SZ3, which replaces linear regression with a dynamic spline interpolation approach. SZ3 can identify nonlinear patterns, resulting in a high compression ratio and superior data quality (e.g., PSNR, etc.). 
Based on the related work review the state-of-the-art calls for new lossy compression methods that can remarkably reduce the size of latent states in comparison to conventional autoencoder-based compressors. With this objective in mind, the following section introduces a prediction-quantization-entropy coder paradigm and presents Deep Dict as an innovative lossy compressor designed to improve prediction capabilities.  

\section{Methodology}
\label{sec: method}

This section lays out the technical details of the proposed Deep Dict framework for lossy time series compression. 

\subsection{Problem Definition}
\label{sec: problem def}
Time series are described as a collection of time-dependent data that can be categorized broadly into univariate time series (UTS) and multivariate time series (MTS).
AE-based compressors encode time series into a latent representation consisting of floating-point numbers; thus, the size of the latent representation has a direct effect on the compression ratio. 

\subsection{Deep Dict}
\label{sec: deep dict}

\subsubsection{Overview}

\begin{figure}
    \centering
    \includegraphics[width=1.0\linewidth]{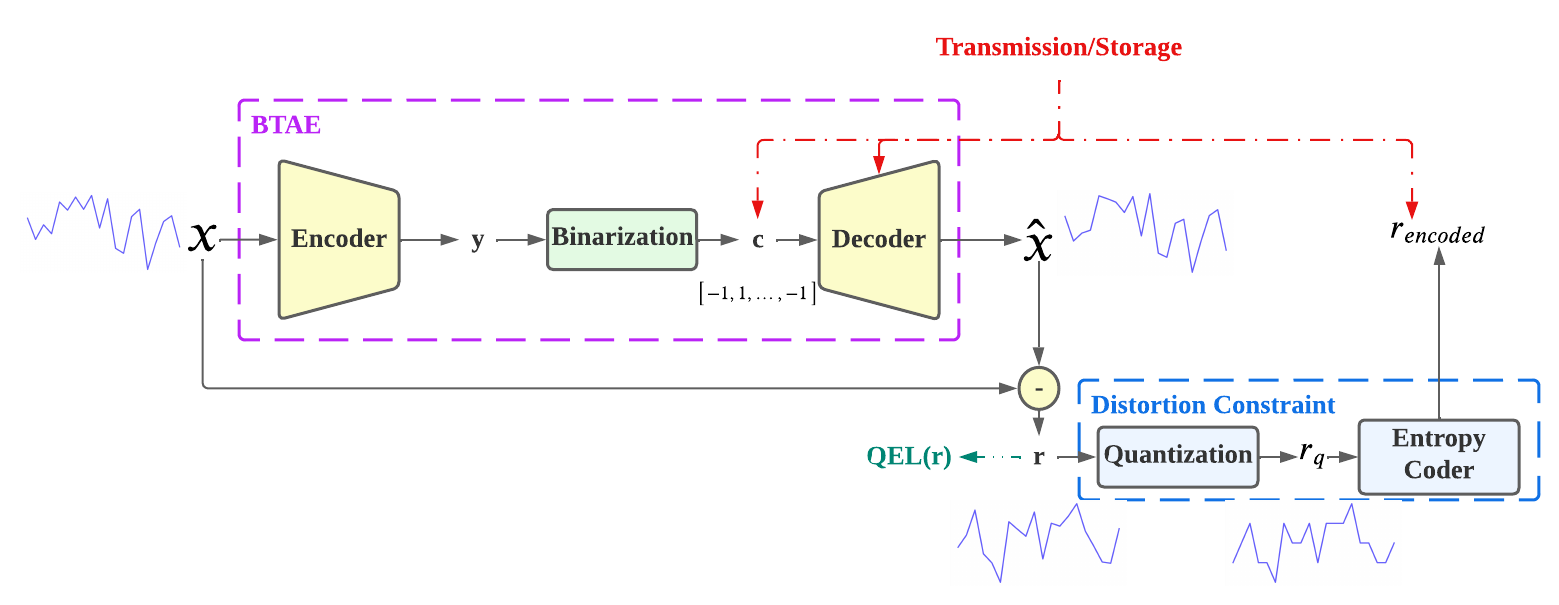}
    \caption{Overview of Deep Dict.}
    \label{fig:deep dict}
\end{figure}

\begin{figure*}
    \centering
    \includegraphics[width=0.75\linewidth]{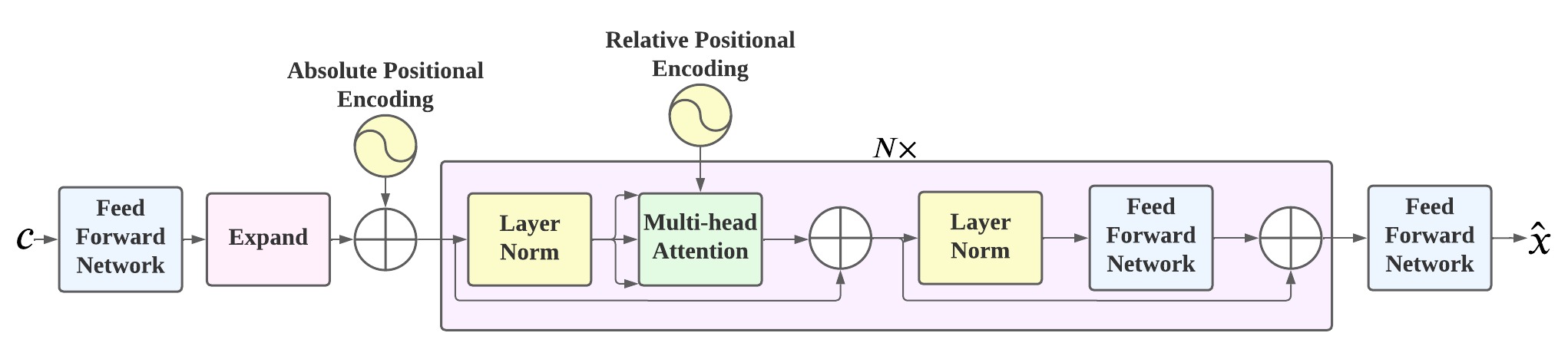}
    \caption{
     Detailed Architecture of the Decoder of Deep Dict.
    }
    \label{fig:decoder}
\end{figure*}

\begin{figure}
    \centering
    \includegraphics[width=0.95\linewidth]{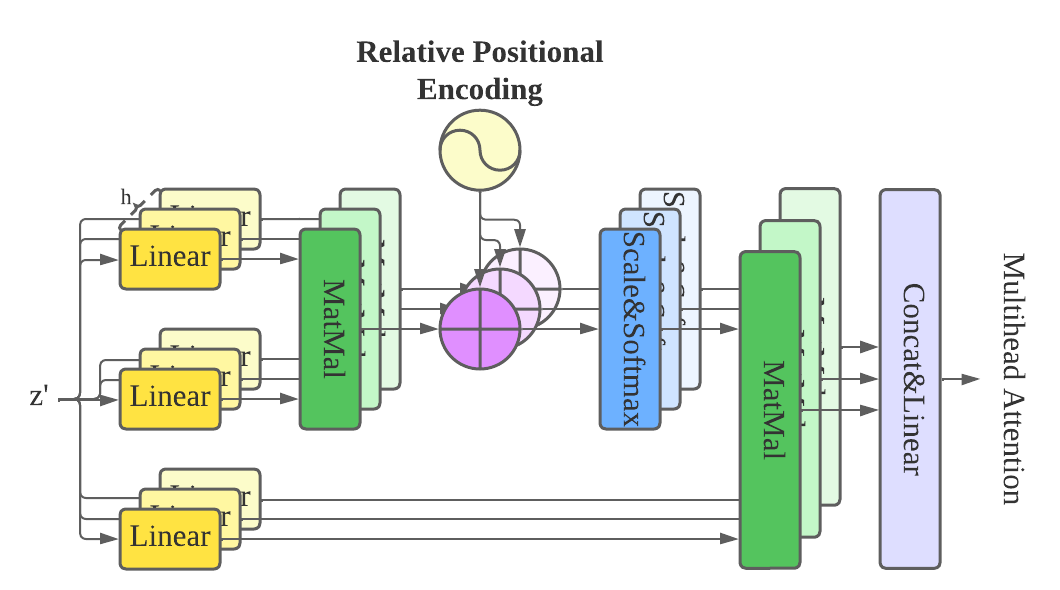}
    \caption{
    Detailed Architecture of Multihead Attention with RPE.
        }
    \label{fig:mha}
\end{figure}

The two major components of Deep Dict are the BTAE and distortion constraint. Figure \ref{fig:deep dict} illustrates an overview of the proposed Deep Dict compressor. Initially, Deep Dict chunks the original long time series into smaller time series $x$ by a time window. BTAE encodes $x$ into Bernoulli latent states $c$ and decodes $c$ to predict time series $\hat{x}$. In order to limit the error of the reconstructed time series to a desired range, the residual $r=x-\hat{x}$ is quantized uniformly to $r_q$ and an entropy coder is used to compress $r_{encoded}$ in a lossless manner. For transmission or storage after compression, $c$, the decoder, and $r_{encoded}$ are used. During decompression, $c$ is fed into the decoder to recover $\hat{x}$, and the entropy coder decodes $r_{encoded}$ to $r_q$.
Since $c$ contains limited information, a feed-forward network (FFN) is used as the encoder in this study.
Fig. \ref{fig:decoder} demonstrates the architecture of the decoder in detail. Due to the fact that each input time series $x$ is truncated from a long sequence, the relative position can be more meaningful than the absolute position.
Therefore, we include relative positional encoding in the proposed decoder. Figure \ref{fig:mha} illustrates the details of multihead attention (MHA) with RPE \cite{huang_music_2018}. 

\subsubsection{Distortion Constraint}

\begin{figure}
    \centering
    \includegraphics[width=0.92\linewidth]{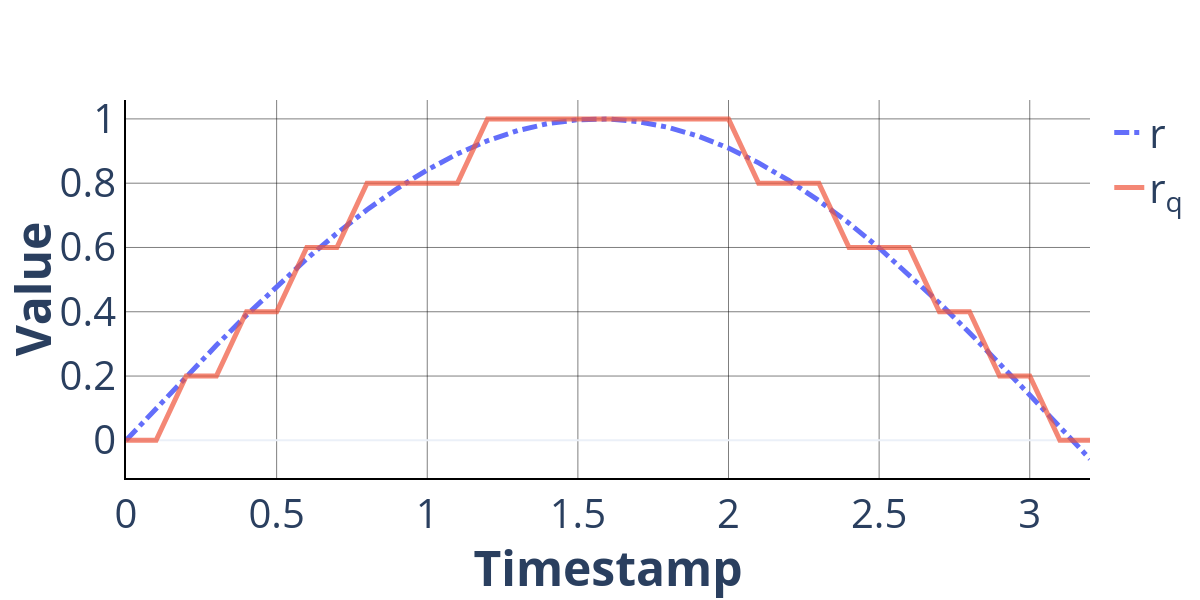}
    \caption{
    Intuitive Example of Uniformed Quantization.
    }
    \label{fig:r-r_q}
\end{figure}


The predicted time series $\hat{x}$ typically has more than $10\%$ mean absolute percentage error (MAPE) loss. By utilizing BTAE, distortion can be constrained to a small range at the expense of more parameters. Thus, with a limited number of parameters in BTAE, the distortion constraint is utilized to reduce the distortion to a desired range. As depicted in Fig. \ref{fig:r-r_q}, $r$ is quantized uniformly to $r_q$ as follows: $r_q=2\epsilon \times round(r / 2\epsilon)$, where $\epsilon$ is the desired MaAE. To avoid float-point overflow, $r_q$ is stored in 64-bit format.
In this work, we utilize adaptive quantized local frequency coding, powered by libbsc, a renowned library for lossless compression \cite{grebnov_ilyagrebnovlibbsc_2022}, as the entropy coder to encode $r_q$ into $r_{encoded}$.

\subsubsection{Quanized Entropy Loss (QEL)}
We introduce a new loss function, QEL, so as to minimize the size of $r_{encoded}$. Good-performing entropy coders with high compression ratios tend to approach the limits,
\begin{equation}
\label{eq:size limit}
    |r_{encoded}| \geq -\sum_{j=0}^{|S|} n(s_j) \log p(s_j) ,
\end{equation}
\noindent where $s_j$ are is the unique values of $r_q$, $n(s_j)$ is a counter to keep track of the number of times $s_j$ appears in $r_q$, and $p(s_j)$ specifies the probability of $s_j$ appearing in $r_q$, respectively.  In light of these, we formulate the minimization problem of the objective function as 
\begin{equation}
\label{eq:H}
    \min_{r} H(r) = -\sum_{j=0}^{|S|} p(s_j) \log p(s_j) .
\end{equation}
Therefore, the minimization process consist of the forward process the QEL calculates the entropy of time series. In backpropagation process we can show that the gradient of $H$ can be writeen as
\begin{equation}
    \label{eq:part H 2}
    \frac{\partial H}{\partial r_i} = 
    \lim_{b \to \infty} \sum_{j=0}^{|S|} [1 + \ln p(s_j)] \times R(r_i - s_j) ,
\end{equation}
\noindent where
\begin{equation}
\label{eq:R}
    R(r_i - s_j) = \frac{b}{|r|\epsilon^b}  \frac{(r_i - s_j)^{b-1}}{[\frac{(r_i - s_j)^b}{\epsilon^b} + 1]^2} .
\end{equation}

\section{Experimental Results}
\label{sec: exp}
BTAE's encoder has 3 layers with 64 hidden states. The FFN that is used for augmenting $c$ has 1 layer with 64 hidden states. The decoder of BTAE has two layers and all feed-forward layers inside the decoder have 64 hidden states. The hyperparameter $d_{model}$ of the transformer encoder is set at 32, and the number of heads of MHA is 8. The FFN used for projecting output time series has 1 layer with 64 hidden states.
We apply the Gaussian Error Linear Unit (GeLU) as the activation function. The following three loss functions are used: L1, L2, and QEL. The other hyperparameter $b$ (for QEL) is set to 10 as default. 
The batch size is set to 64. Adam optimizer is used with a learning rate of 0.0001, weight decay of 0.01, $\beta_1$ of 0.9, and $\beta_2$ of 0.999.
The model is trained by using PyTorch 1.11 on NVIDIA GeForece RTX 3070 and Intel Xeon W-2295.

\begin{table*}[]
\vspace{0.03in}
\centering
\caption{
Compression ratio compared to the best baseline methods.
}
\label{tab:uts comparision}
\resizebox{\linewidth}{!}{%
\begin{tabular}{l||l|llll|ll|ll|ll}
\hline
Dataset &
  length &
  CA &
  SZ &
  LFZip &
  SZ3 &
  DeepDict(L1) &
  Imp. &
  DeepDict(QEL) &
  Imp. &
  DeepDict(RPE) &
  Imp. \\ \hline \hline
dna &
  1167877 &
  4.86 &
  \textbf{8.62} &
  8.40 &
  7.78 &
  8.58 &
  -0.46\% &
  8.09 &
  -6.15\% &
  {\color[HTML]{ED7D31} 8.36} &
  {\color[HTML]{ED7D31} -3.02\%} \\ 
pow &
  2049280 &
  12.47 &
  23.99 &
  17.98 &
  \textbf{24.21} &
  23.92 &
  -1.20\% &
  {\color[HTML]{70AD47} 23.98} &
  {\color[HTML]{70AD47} -0.95\%} &
  {\color[HTML]{ED7D31} 24.00} &
  {\color[HTML]{ED7D31} -0.87\%} \\ 
watch\_gyr &
  3205431 &
  10.75 &
  24.79 &
  \textbf{28.77} &
  26.85 &
  27.10 &
  -5.80\% &
  24.68 &
  -14.22\% &
  {\color[HTML]{ED7D31} 25.63} &
  {\color[HTML]{ED7D31} -10.91\%} \\ 
watch\_acc &
  3540962 &
  5.19 &
  11.00 &
  12.71 &
  10.78 &
  {\color[HTML]{FF0000} \textbf{13.24}} &
  {\color[HTML]{FF0000} \textbf{4.17\%}} &
  \textbf{12.74} &
  \textbf{0.24\%} &
  {\color[HTML]{ED7D31} \textbf{12.87}} &
  {\color[HTML]{ED7D31} \textbf{1.26\%}} \\ 
phones\_acc &
  13062475 &
  7.13 &
  12.63 &
  14.12 &
  12.64 &
  {\color[HTML]{FF0000} \textbf{15.95}} &
  {\color[HTML]{FF0000} \textbf{12.96\%}} &
  \textbf{15.84} &
  \textbf{12.18\%} &
  {\color[HTML]{ED7D31} \textbf{15.89}} &
  {\color[HTML]{ED7D31} \textbf{12.54\%}} \\ 
phones\_gyr &
  13932632 &
  27.32 &
  52.00 &
  46.28 &
  55.11 &
  {\color[HTML]{FF0000} \textbf{56.44}} &
  {\color[HTML]{FF0000} \textbf{2.41\%}} &
  \textbf{55.46} &
  \textbf{0.64\%} &
  52.66 &
  -4.45\% \\ 
bar\_crawl &
  14057567 &
  4.99 &
  18.09 &
  19.14 &
  18.39 &
  \textbf{27.57} &
  \textbf{44.04\%} &
  {\color[HTML]{70AD47} \textbf{29.41}} &
  {\color[HTML]{70AD47} \textbf{53.66\%}} &
  {\color[HTML]{ED7D31} \textbf{29.78}} &
  {\color[HTML]{ED7D31} \textbf{55.59\%}} \\ 
soybeans &
  22824499 &
  3.43 &
  14.35 &
  15.61 &
  13.50 &
  \textbf{17.04} &
  \textbf{9.16\%} &
  {\color[HTML]{70AD47} \textbf{18.32}} &
  {\color[HTML]{70AD47} \textbf{17.36\%}} &
  {\color[HTML]{ED7D31} \textbf{18.45}} &
  {\color[HTML]{ED7D31} \textbf{18.19\%}} \\ 
synthetic &
  43000000 &
  113.07 &
  119.00 &
  58.35 &
  127.23 &
  \textbf{125.16} &
  \textbf{-1.63\%} &
  {\color[HTML]{70AD47} \textbf{154.65}} &
  {\color[HTML]{70AD47} \textbf{21.55\%}} &
  \textbf{149.54} &
  \textbf{17.54\%} \\ 
ppg\_ecg &
  90642300 &
  46.01 &
  69.20 &
  45.47 &
  71.42 &
  \textbf{90.31} &
  \textbf{26.45\%} &
  {\color[HTML]{70AD47} \textbf{95.31}} &
  {\color[HTML]{70AD47} \textbf{33.45\%}} &
  {\color[HTML]{ED7D31} \textbf{97.28}} &
  {\color[HTML]{ED7D31} \textbf{36.21\%}} \\ \hline
\end{tabular}%
}
\end{table*}

\begin{figure}
    \centering
    \includegraphics[width=0.95\linewidth]{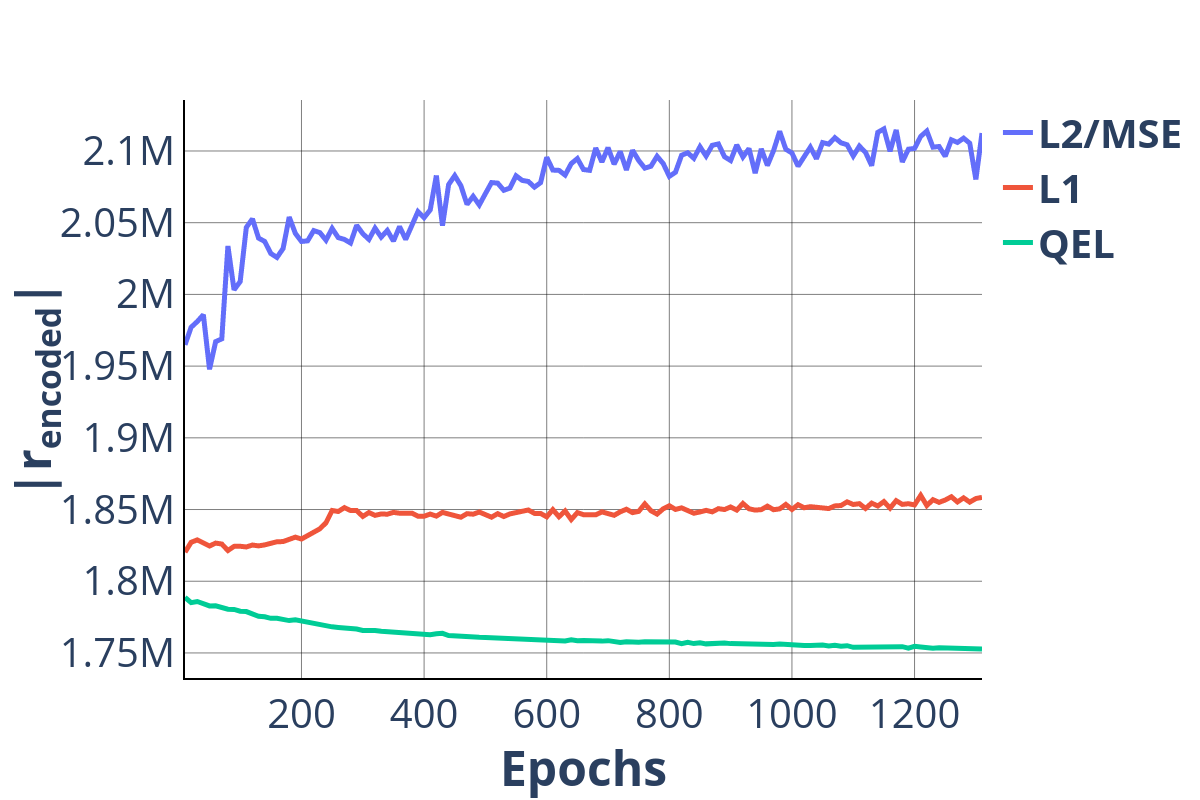}
    \caption{
    Comparison of L1, L2/MSE, and QEL under bar\_crawl univariate dataset.
    }
    \label{fig:l1 qel}
\end{figure}
\begin{figure*}
    \centering
    \includegraphics[width=0.75\linewidth]{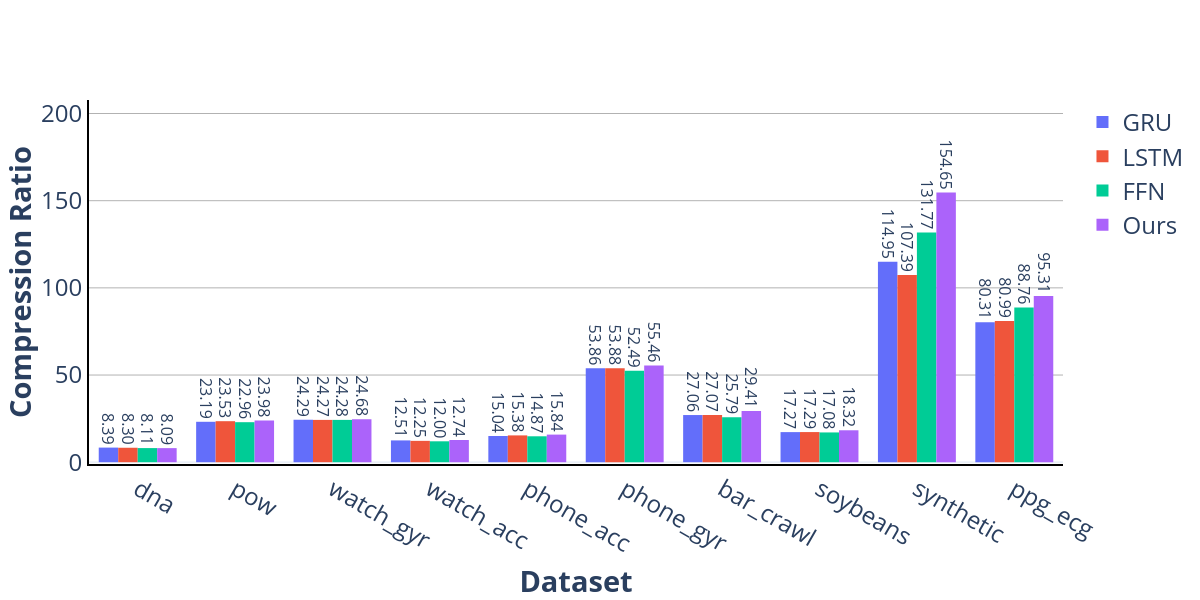}
    \caption{Comparison among different architectures of the decoder on univariate datasets.}
    \label{fig:cr vs decoder}
\end{figure*}

\begin{table*}[]
\centering
\caption{Compression ratio comparison between univariate and multivariate mode.\iffalse under 0.1 MaAE. Uni.: Univariate mode; Mul.: Multivariate mode; Imp.: The improvement of multivariate mode compared to univariate mode \fi}
\label{tab:mts comparison}
\resizebox{\linewidth}{!}{%
\begin{tabular}{l||l|lll|lll|lll}
\hline
\multicolumn{1}{c||}{{}} &
  \multicolumn{1}{c|}{{}} &
  \multicolumn{3}{c|}{{L1}} &
  \multicolumn{3}{c|}{{QEL}} &
  \multicolumn{3}{c}{{RPE}} \\ \cline{3-11} 
\multicolumn{1}{c||}{\multirow{-2}{*}{{Dataset}}} &
  \multicolumn{1}{c|}{\multirow{-2}{*}{{Shape}}} &
  {Uni.} &
  {Mul.} &
  {Imp.} &
  {Uni.} &
  {Mul.} &
  {Imp.} &
  {Uni.} &
  {Mul.} &
  {Imp.} \\ \hline \hline
{watch\_gyr} &
  {3205431 $\times$ 3} &
  {\textbf{31.14}} &
  {28.83} &
  {-7.42\%} &
  {\textbf{31.64}} &
  {28.52} &
  {-9.86\%} &
  {\textbf{32.93}} &
  {28.72} &
  {-12.78\%} \\
{watch\_acc} &
  {3,540,962 $\times$ 3} &
  {\textbf{13.06}} &
  {11.28} &
  {-13.63\%} &
  {\textbf{13.53}} &
  {11.48} &
  {-15.15\%} &
  {\textbf{13.57}} &
  {10.79} &
  {-20.49\%} \\
{phone\_acc} &
  {13062475 $\times$ 3} &
  {11.34} &
  {\textbf{13.15}} &
  {15.96\%} &
  {11.77} &
  {\textbf{13.55}} &
  {15.12\%} &
  {11.8} &
  {\textbf{13.89}} &
  {17.71\%} \\
{phone\_gyr} &
  {13,932,632 $\times$ 3} &
  {37.28} &
  {\textbf{47.09}} &
  {26.31\%} &
  {42.85} &
  {\textbf{47.81}} &
  {11.58\%} &
  {42.1} &
  {\textbf{46.9}} &
  {11.40\%} \\
{bar\_crawl} &
  {14,057,567 $\times$ 3} &
  {28.08} &
  {\textbf{29.1}} &
  {3.63\%} &
  {22.56 (b=3)} &
  {\textbf{28.57} (b=3)} &
  {26.64\% (b=3)} &
  {23.65 (b=3)} &
  {\textbf{28.85} (b=3)} &
  {21.99\% (b=3)} \\
{synthetic} &
  {43,000,000$\times$5} &
  {24.42} &
  {\textbf{42.68}} &
  {74.77\%} &
  {15.9} &
  {\textbf{43.5}} &
  {173.58\%} &
  {28.17} &
  {\textbf{44.31}} &
  {57.29\%} \\ \hline
\end{tabular}%
}
\end{table*}

\begin{table}[]
\caption{
Transferability; between TL and NTL in terms of compression ratio. 
}
\label{tab:transferability}
    \begin{subtable}[t]{\linewidth}
    \centering
    \caption{Univariate datasets. \iffalse Compression ratio comparison between non-transfer learning and transfer learning under univariate datasets\fi}
    \label{tab:trans uni}
    \resizebox{\columnwidth}{!}{%
    \begin{tabular}{l||l|l|ll|ll}
    \hline
    Dataset     & Length   & NTL    & TL     & Imp.     & TL+RPE & Imp.     \\ \hline \hline
    dna         & 1,167,877  & 8.09   & 8.01   & -0.99\%  & 8.02   & -0.87\%  \\
    pow         & 2,049,280  & 23.98  & 21.84  & -8.92\%  & 22.01  & -8.22\%  \\
    watch\_gyr  & 3,205,431  & 24.68  & 24.01  & -2.71\%  & 24.19  & -1.99\%  \\
    watch\_acc  & 3,540,962  & 12.74  & 12.78  & 0.31\%   & 12.73  & -0.08\%  \\
    phones\_acc & 13,062,475 & 15.84  & 14.87  & -6.12\%  & 14.83  & -6.38\%  \\
    phones\_gyr & 13,932,632 & 55.46  & 53.43  & -3.66\%  & 53.76  & -3.07\%  \\
    bar\_crawl  & 14,057,567 & 29.41  & 28.54  & -2.96\%  & 28.71  & -2.38\%  \\
    soybeans    & 22,824,499 & 18.32  & 17.59  & -3.98\%  & 17.85  & -2.57\%  \\
    synthetic   & 43,000,000 & 154.65 & 119.97 & -22.42\% & 120    & -22.41\% \\
    ppg\_ecg    & 90,642,300 & 95.31  & 94.15  & -1.22\%  & 96.67  & 1.43\%   \\ \hline
    \end{tabular}%
    }
    \end{subtable}
    \hfill
    \begin{subtable}[t]{\linewidth}
    \centering
    \caption{Multivariate datasets. \iffalse Compression ratio comparison between non-transfer learning and transfer learning on multivariate datasets\fi}
    \label{tab:trans multi}
    \resizebox{\columnwidth}{!}{%
    \begin{tabular}{l||l|l|ll|ll}
    \hline
    Dataset    & Shape      & NTL   & TL    & Imp.     & TL+RPE & Imp.    \\ \hline \hline
    watch\_gyr & 3,205,431 $\times$ 3  & 28.52 & 21.79 & -23.60\% & 27.19  & -4.66\% \\
    watch\_acc & 3,540,962 $\times$ 3  & 11.48 & 10.92 & -4.88\%  & 10.71  & -6.71\% \\
    phone\_acc & 13,062,475 $\times$ 3 & 13.55 & 12.31 & -9.15\%  & 12.52  & -7.60\% \\
    phone\_gyr & 13,932,632 $\times$ 3 & 47.81 & 43.4  & -9.22\%  & 46.31  & -3.14\% \\
    bar\_crawl & 14,057,567 $\times$ 3 & 28.57 & 28.36 & -0.74\%  & 30.58  & 7.04\%  \\
    synthetic  & 43,000,000 $\times$ 5 & 43.5  & 44.84 & 3.08\%   & 44.13  & 1.45\%  \\ \hline
    \end{tabular}%
    }
    \end{subtable}
\end{table}

\subsection{Numerical Results}
This section discusses Deep Dict's performance under a variety of time series datasets.
\subsubsection{Results on Univariate Datasets}
To evaluate the performance of Deep Dict, five lossy time series compressors are used as baselines. These compressors include Critical Aperture (CA)\footnote{\url{https://github.com/shubhamchandak94/LFZip/blob/master/src/ca_compress.py}}, SZ2\footnote{\url{https://github.com/szcompressor/SZ}}, LFZip\footnote{\url{https://github.com/shubhamchandak94/LFZip}}, and SZ3\footnote{\url{https://github.com/szcompressor/SZ3}}. CA is an industrially well-received compressor that is computationally simple and efficient. 

Table \ref{tab:uts comparision} compares the proposed method (RPE is not used by default) to the baselines under the datasets that are ordered with respect to the length of their time series. Under seven out of ten datasets, the proposed method outperforms the state-of-the-art algorithms. Due to the overhead of BTAE and codes, Deep Dict performs similarly to the baseline on small datasets; however, under large datasets, Deep Dict outperforms the baselines by at most $53.66\%$. 


As depicted in Fig.  \ref{fig:l1 qel}, when the bar\_crawl dataset is considered as a representative example, because L1 and L2 are not particularly designed to reduce the size of $r_{encoded}$, L1 and L2 losses can result in an increase of $|r_{encoded}|$ during the training process; however, QEL can handle such situations and increase the compression ratio. We further enable the RPL (using QEL loss) in Deep Dict. The results indicate that RPL can improve the performance of Deep Dict on eight out of ten datasets. 

As demonstrated in Fig.  \ref{fig:deep dict}, the decoder of BTAE can be other network architecture intended for time series data, such as FFN, LSTM, or GRU. In order to illustrate the effectiveness of the proposed decoder, we compare the various decoder designs in Fig.  \ref{fig:cr vs decoder}. Similar to the typical decoder of an RNN-based autoencoder, the auto-regressive approach is employed for LSTM and GRU, with $c$ serving as the initial input timestamp. Under six out of ten datasets, the results indicate that LSTM performs better than FFN and GRU. Our proposed decoder outperforms the other network architectures under nine out of ten datasets. 

\subsubsection{Results on Multivariate Datasets}
Table \ref{tab:mts comparison} compares the performance of the univariate mode (i.e., flattening the MTS prior to feeding into Deep Dict) and multivariate mode. Compared with the performance of L1 and QEL, QEL outperforms L1 on all datasets except for the bar\_crawl dataset. When RPL is leveraged (QEL is used as a loss function), it is able to improve the performance of the univariate and multivariate modes.

\subsubsection{Transferability}
\label{sec:transferability}
As shown in Table \ref{tab:trans uni}, the compression ratio of Deep Dict with transfer learning (Deep Dict + TL) reduces by less than 5\% under 7 out of 10 univariate datasets when compared to training a model from scratch. Under five out of seven univariate datasets (where Deep Dict outperforms the best baseline), Deep Dict + TL continues to outperform the best baseline. Table \ref{tab:trans multi} shows the comparative results between NTL and TL for multivariate datasets. Under five out of six multivariate datasets, Deep Dict + TL decreases the compression ratio by up to $9.22\%$.

\subsubsection{Empirical Study}

\label{sec:empirical}
\begin{figure}
    \centering
    \includegraphics[width=0.95\linewidth]{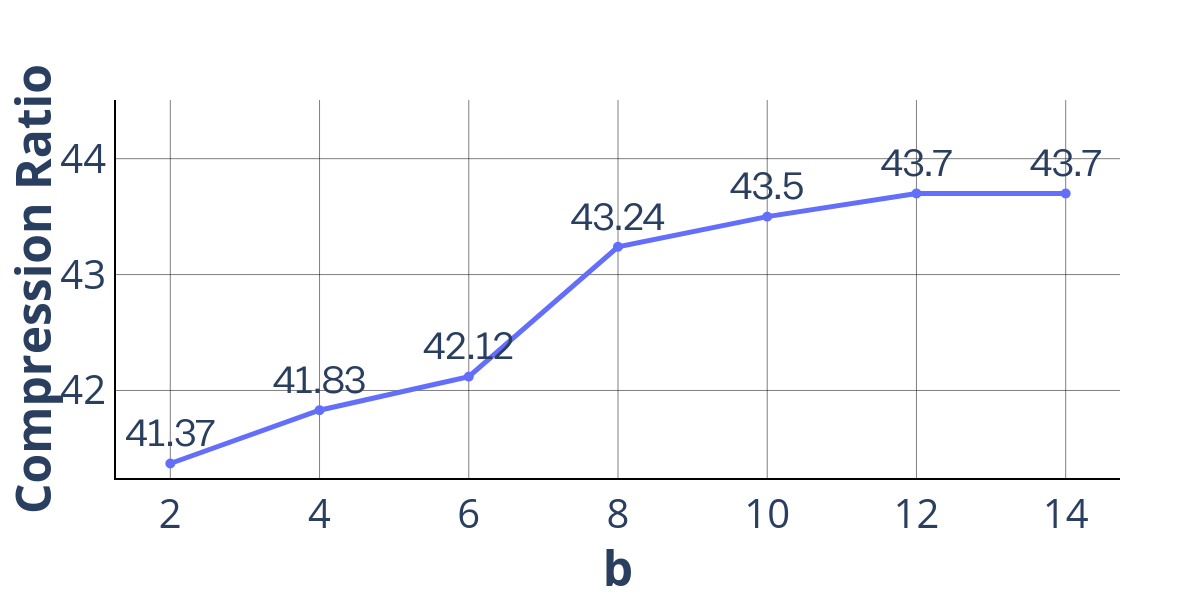}
    \caption{The effect of $b$ on compression ratio. \iffalse The effect of $b$ on compression ratio; synthetic dataset (43000000 $\times$ 5) is used. \fi}
    \label{fig:cr vs b}
\end{figure}
As shown in Fig.  \ref{fig:cr vs b}, the compression ratio increases with $b$. When $b > 6$, QEL performs better than L1 loss. 

\begin{figure}
    \centering
    \includegraphics[width=0.95\linewidth]{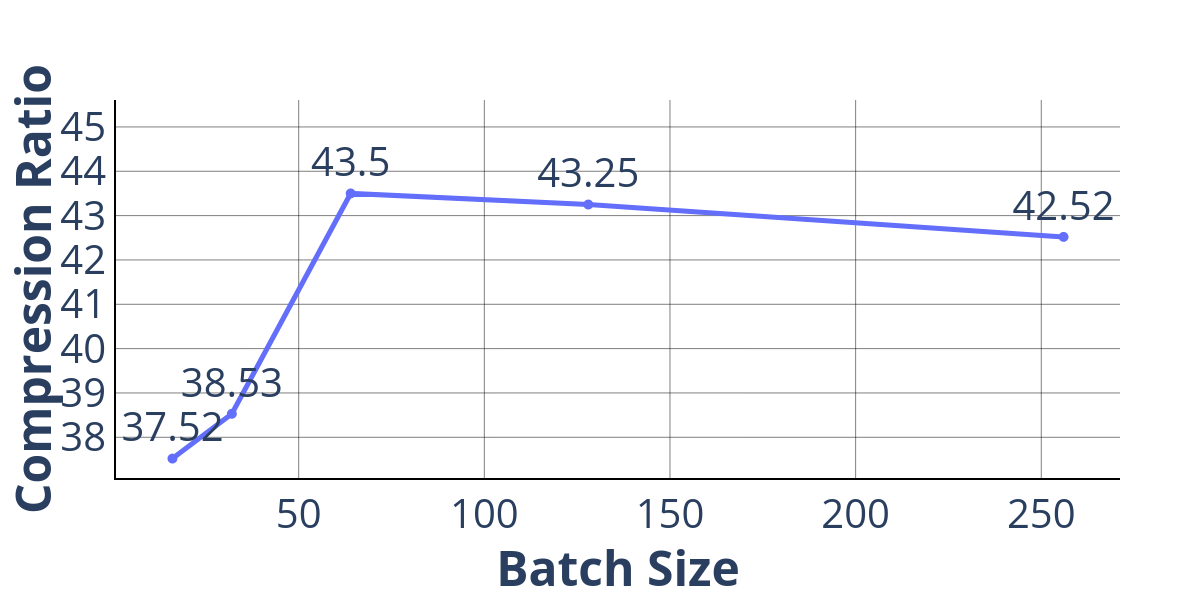}
    \caption{The effect of batch size on compression ratio.\iffalse ; synthetic dataset (43000000*5) is used.\fi}
    \label{fig:cr vs batch size}
\end{figure}
\begin{figure}
    \centering
    \includegraphics[width=0.95\linewidth]{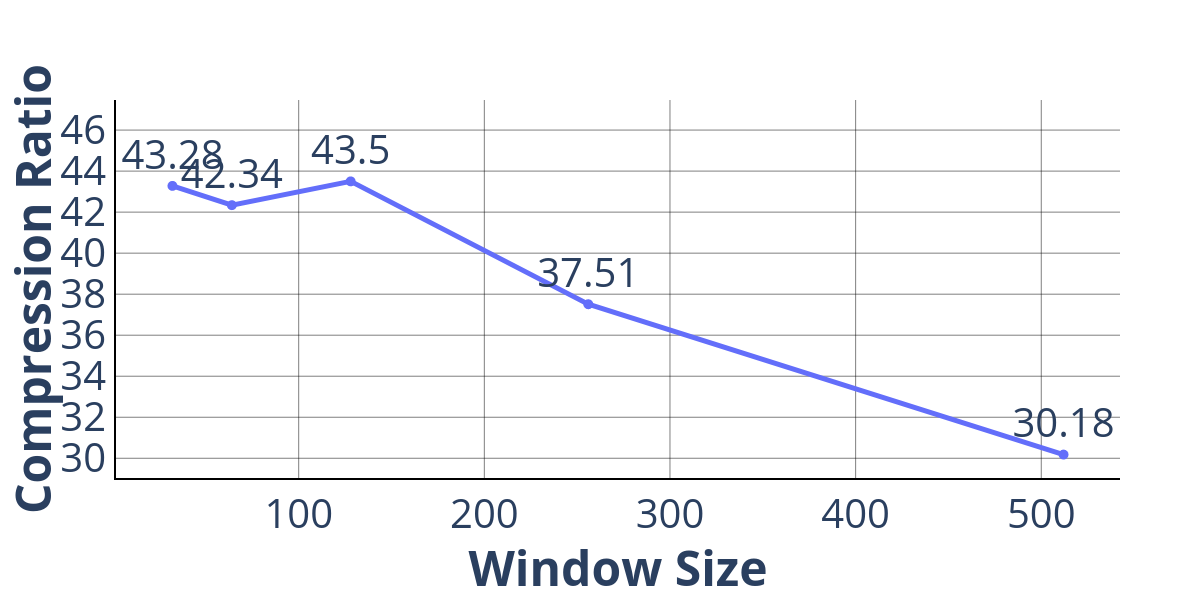}
    \caption{The effect of window size on compression ratio. \iffalse The effect of window size on compression ratio; synthetic dataset (43000000 $\times$ 5) is used.\fi}
    \label{fig:cr vs window size}
\end{figure}
Since QEL can only minimize the entropy of each batch for each backpropagation, batch size is one of the critical hyperparameters for QEL. Figure \ref{fig:cr vs batch size} indicates that Deep Dict achieves the highest compression ratio when the batch size is 64, and that compression ratio is steady for batch sizes greater than 64. 
As seen in Fig. 
\ref{fig:cr vs window size}, Deep Dict performs effectively with a small window, however, its compression ratio decreases rapidly under a large window.

\begin{figure}
    \centering
    \includegraphics[width=0.95\linewidth]{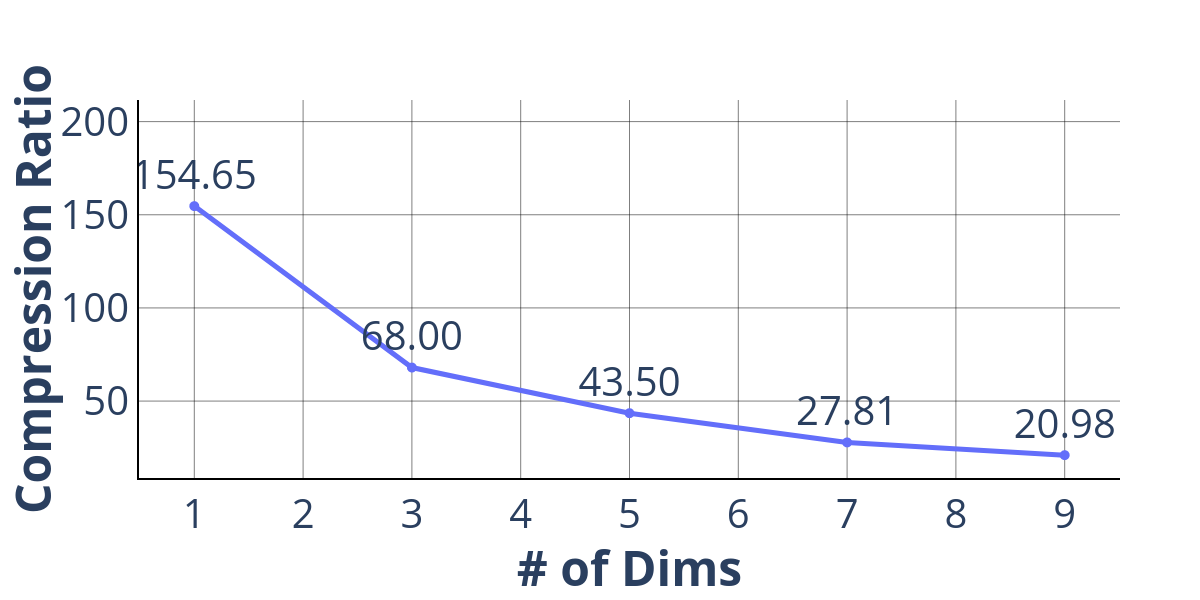}
    \caption{The change of compression ratio with the number of dimension of data. \iffalse ; synthetic datasets are used which has the same length (43000000).\fi}
    \label{fig:cr vs dim}
\end{figure}
\begin{figure}
    \centering
    \includegraphics[width=0.95\linewidth]{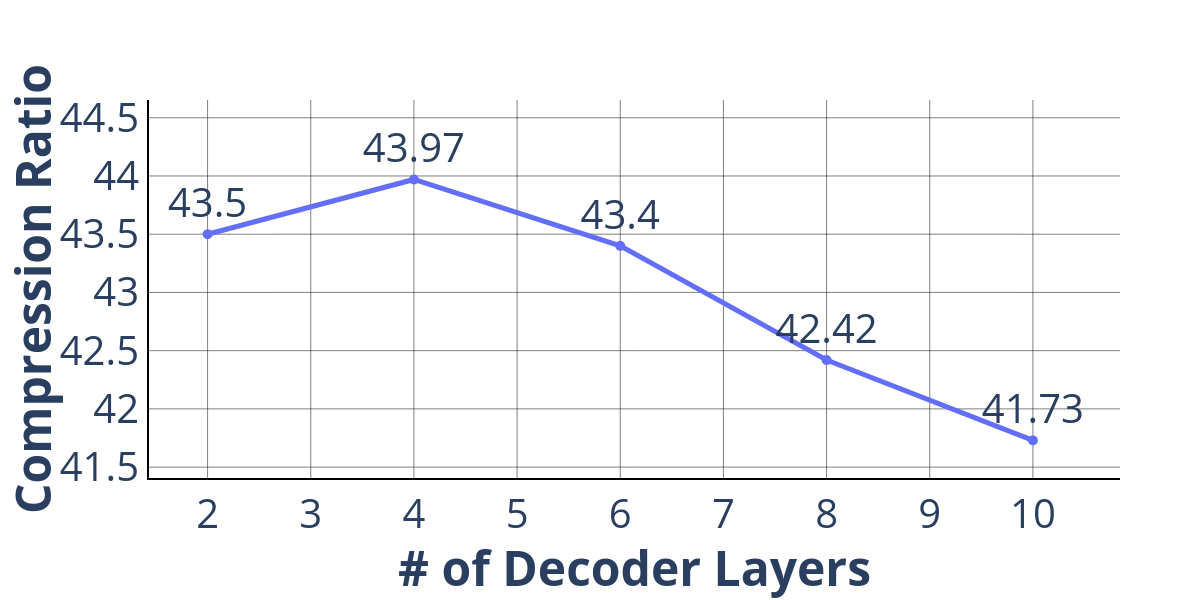}
    \caption{The effect of the number of layers on compression ratio.\iffalse ; synthetic dataset (43000000 $\times$ 5) is used.\fi}
    \label{fig:cr vs layers}
\end{figure}
\begin{figure}
    \centering
    \includegraphics[width=0.95\linewidth]{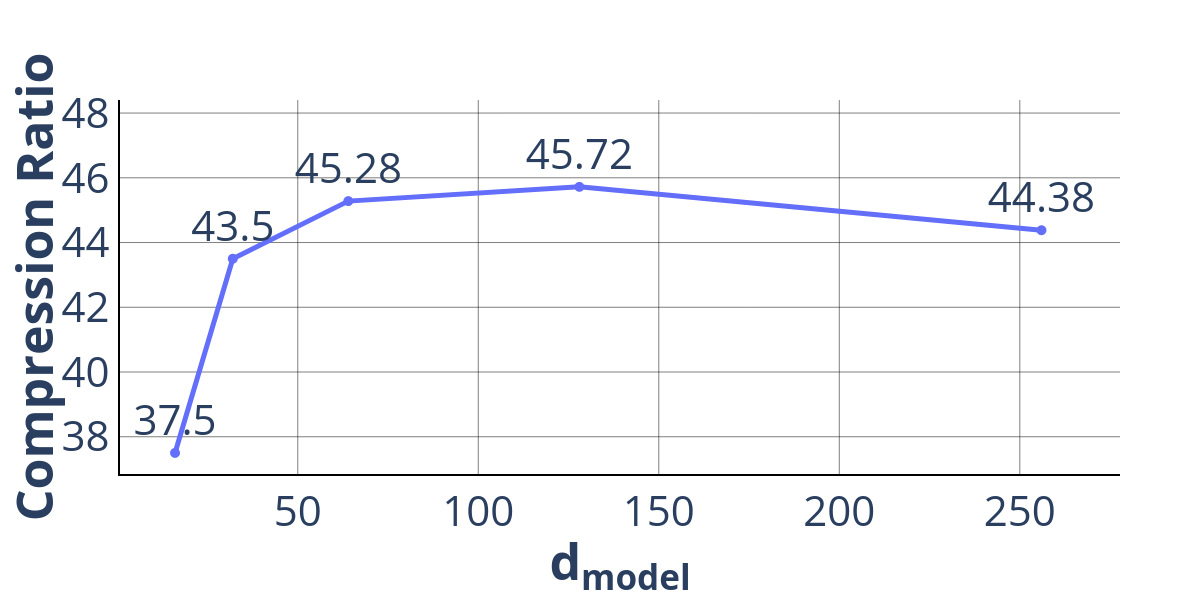}
    \caption{The effect of $d_{model}$ on compression ratio. \iffalse ; synthetic dataset (43000000 $\times$ 5) is used.\fi}
    \label{fig:cr vs d_model}
\end{figure}
\begin{figure}
    \centering
    \includegraphics[width=0.95\linewidth]{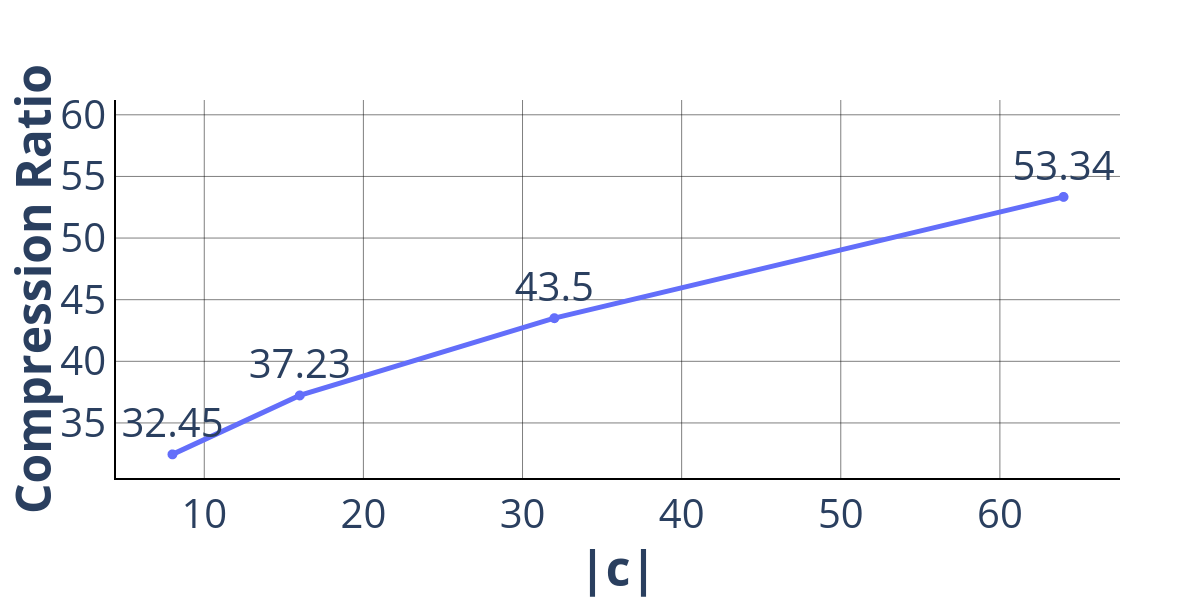}
    \caption{The effect of $|c|$ on compression ratio. \iffalse ; synthetic dataset (43000000 $\times$ 5) is used. \fi}
    \label{fig:cr vs c}
\end{figure}
Previous results indicate that Deep Dict outperforms the baselines under large time series datasets. Figure \ref{fig:cr vs dim} illustrates the effect of the dimensionality on compression ratio (with the same hyperparameters). As network size cannot be increased, Deep Dict's compression ratio is limited by the number of parameters. There are two ways to increase the number of parameters: stacking more layers and expanding the network. As shown in Fig.  \ref{fig:cr vs layers}, stacking more transformer encoders does not result in a significant improvement; rather, as the number of layers increases, the compression ratio decreases because of the increase in the decoder size. On the other hand, Fig.  \ref{fig:cr vs d_model} demonstrates that compression ratio can be improved with a large $d_{model}$ (one of the Transformer hyperparameters). It is worth noting that large $d_{model}$ is not suitable for small datasets since large $d_{model}$ will notably increase the number of parameters in BTAE. Figure \ref{fig:cr vs c} depicts the variation of compression ratio under varying Bernoulli latent states ($|c|$). Increasing $|c|$ is possible to considerably enhance Deep Dict performance for big datasets, although, similar to $d_{model}$, a large $|c|$ can also increase the number of parameters. In summary, increasing $b$, $d_{model}$, and $|c|$ can further enhance the performance of long-time series.

\section{Conclusion}
\label{sec: conclusion}
We propose a Bernoulli transformer autoencoder-based lossy time series compressor, namely Deep Dict to improve the compression ratio by learning the Bernoulli representation of time series. We have substituted the conventional regression loss with a novel loss function, quantized entropy loss (QEL), which further improves the compression ratio and reduces the difficulty of optimization. Deep Dict outperforms the state-of-the-art time series compression under $7$ out of 10 datasets, particularly under the lengthy time series datasets. We have shown that Deep Dict can outperform the best baseline by a maximum of $53.66\%$. The proposed loss function, QEL can boost compression ratios more than the traditional regression losses such as L1 and L2. The experiment on the transferability demonstrates that Deep Dict can be accelerated by transfer learning without significantly sacrificing much compression ratio (less than $5\%$). Moreover, RPE can improve Deep Dict's transferability on multivariate datasets. When multivariate and univariate modes are compared, the results indicate that Deep Dict's multivariate mode performs better under larger multivariate time series. 
In our future work, we are focusing on utilizing neural network quantization to reduce the size of the model further. Furthermore, as various data sizes and types have different hyperparameters, selecting hyperparameters automatically based on the data is also on our agenda.

\section*{Acknowledgment}
This work was supported in part by the Natural Sciences and Engineering Research Council of Canada (NSERC) under Grant RGPIN/2017-04032. Petar Djukic was with Ciena (Kanata, ON, Canada) when this work was done.

\bibliographystyle{IEEEtran}

\end{document}